\documentclass[floatfix,aps,prb,reprint,superscriptaddress]{revtex4-2}
\usepackage[utf8]{inputenc}
\usepackage{amsmath,amssymb,amsfonts}
\usepackage{graphicx}
\usepackage{bm}
\usepackage{color}
\usepackage{booktabs}
\usepackage{hyperref}
\hypersetup{colorlinks=true,linkcolor=blue,citecolor=blue,urlcolor=blue}
\newcommand{\dd}{\mathrm d}

\begin{document}

\title{Nonperturbative renormalization of Haldane pseudopotentials from the exact two-electron spectrum}

\author{G.-Q. Hai}
\email{email: hai@ifsc.usp.br}
\author{M. T. Matsubara}
\affiliation{Instituto de F\'{i}sica de S\~{a}o Carlos, Universidade de S\~{a}o Paulo, 13560-970, S\~{a}o Carlos, SP, Brazil}
\author{L. C\^andido}
\affiliation{Instituto de F\'{i}sica,  Universidade Federal de Goi\'{a}s, 74.001-970, Goi\^{a}nia, GO, Brazil}
\author{B. G. A. Brito}
\affiliation{Departamento de F\'{i}sica, Universidade Federal do Tri\^{a}ngulo Mineiro, 38.025-180, Uberaba, MG, Brazil}

\begin{abstract}
We investigate how exact two-electron energies and radial wavefunctions can improve many-electron descriptions of
Landau-level (LL) mixing. Renormalized pseudopotentials are defined from the lowest interacting relative-motion
branch in each lowest-LL angular-momentum channel. They include changes in Coulomb and magnetic kinetic expectation
values nonperturbatively and recover the leading two-body perturbative correction. Matched three- and four-electron
benchmarks show that the scalar renormalized model improves lowest fixed-sector energies and first fixed-sector
spacings over conventional lowest-LL projection, although the tested perturbative models give more accurate lowest energies.
Retaining the same second-order induced three-body term while replacing the perturbative two-body input by
the renormalized pseudopotentials yields a hybrid with smaller absolute lowest-energy errors throughout the sampled
coupling range. At the Coulomb-to-cyclotron energy ratio $\gamma_B=3.5$, the errors decrease by approximately $58\%$
and $57\%$ for three and four electrons, respectively, relative to the tail-completed two-plus-three-body perturbative
comparator. At the same coupling, complementary antisymmetric multi-LL variational states built from the pair radial
amplitudes recover $99.41\%$ and $99.11\%$ of the matched finite-cutoff LL-mixing energy lowering for five and six
electrons, outperforming equal-rank first-order-pair controls. These results establish pair energies and radial structure
as complementary, quantitatively useful inputs for many-electron treatments of LL mixing.
\end{abstract}
\maketitle

\section{Introduction}\label{sec:intro}
Electron correlations within a Landau level are central to the fractional quantum Hall effect\cite{Tsui1982,Laughlin1983,Jain1989}.
Haldane pseudopotentials provide a natural description of rotationally invariant two-body interactions by resolving them into
relative-angular-momentum channels\cite{Haldane1983}. Their usefulness extends beyond the two-electron problem: the same pair
projectors act on all overlapping pairs in an antisymmetric many-electron Hilbert space. When interactions admix higher Landau
levels (LLs), however, the internal radial structure of each pair can relax. This raises the question of how the resulting pair
energies and radial wavefunctions can supply useful input to effective interactions and collective many-electron trial states,
respectively.

LL mixing is conventionally characterized by $\kappa$, the ratio of the Coulomb scale to the cyclotron energy.
For the parabolic-band Coulomb model considered here, this energy ratio $\kappa$ equals $\gamma_B\equiv\ell_B/a_B^*$, the ratio
of the magnetic length to the effective Bohr radius. We primarily use $\gamma_B$ to emphasize this length-scale interpretation
and retain $\kappa$ when connecting with the LL-mixing literature. Their explicit definitions will be given in Sec.~\ref{sec:two}.
Perturbative effective-Hamiltonian treatments generate corrections to two-body pseudopotentials together with induced three-body
interactions at leading nontrivial order\cite{Bishara2009,Peterson2013,Sodemann2013,Simon2013}. Their many-electron consequences
include changes in competing fractional states\cite{Rezayi2017}. Explicit multi-LL methods\cite{Zaletel2015} and fixed-phase
diffusion Monte Carlo\cite{Zhang2016} provide complementary approaches beyond a low-order effective interaction.

When $\kappa$ is of order unity or larger, retaining the radial response beyond a low-order coupling expansion becomes important.
Eliminating higher LLs generates a hierarchy of induced many-body interactions, while explicit multi-LL calculations face
a Hilbert space that grows rapidly with particle number and cutoff. The isolated interacting pair provides a tractable
starting point:
its relative-motion problem can be solved nonperturbatively, retaining radial LL mixing without a coupling expansion. We use
this pair information as input and test its quantitative usefulness in finite many-electron systems under matched conditions.

Analytical solutions of the quantum two-particle problem in a uniform perpendicular magnetic field have been obtained at particular
parameter values\cite{Vercin1991,Taut1994}; the corresponding classical dynamics has also been studied\cite{Curilef1997}. Its
center-of-mass (CM) and relative motions separate exactly, and the Coulomb interaction acts only on the relative coordinate.
The interaction-resolved relative branches and their radial structure are discussed in Ref.~\cite{Hai2024}. Here we use the lowest
interacting relative energy in each lowest-Landau-level (LLL) relative-angular-momentum channel ($m\leq 0$ in our convention)
to define the renormalized pseudopotentials $V^*_{|m|}$ and their corrections $\Delta_{|m|}$ relative to the conventional
Haldane pseudopotentials $V_{|m|}$. Their formal definition is given in Sec.~\ref{sec:two}. Relative to the strictly projected
LLL pair state, the renormalized pseudopotentials include the reduction in the Coulomb expectation value together with the excess
magnetic kinetic energy associated with higher-LL admixture in the stationary correlated state. The kinetic energy within the strictly
projected LLL remains fixed. The corresponding radial functions $h_0^{|m|}$ supply additional information beyond a channel energy.

The present work builds on the interacting relative-motion branches, correlation-resolved radial functions
$h_{\alpha}^{|m|}(x)$, and correlated rotating electron pair (CREP) description introduced in Ref.~\cite{Hai2024}.
The present paper develops four connected applications of this pair information:
(i) the spectral definition of renormalized pseudopotentials $V^*_{|m|}$ and the decomposition of their corrections into
kinetic and Coulomb contributions; (ii) the explicit connection to the leading perturbative two-body correction;
(iii) the embedding of the renormalized coefficients in a scalar many-electron Hamiltonian and in a hybrid retaining
a matched induced three-body term; and
(iv) the use of pair radial amplitudes to generate collective multi-LL variational states. Section~II
summarizes the two-electron formulation needed for these developments.

Our purpose is to assess the quantitative usefulness of these two forms of pair input under matched many-electron conditions.
For three and four fully spin-polarized electrons, we compare conventional LLL projection, a scalar Hamiltonian using $V^*_{|m|}$,
second-order effective Hamiltonians, and physical multi-LL benchmarks. A hybrid construction retains the same perturbative
three-body term as the second-order comparator while replacing its two-body coefficients by the renormalized pseudopotentials.
This comparison tests the pair input at a common approximation for the induced many-body contribution.

The comparisons identify both a useful application and the limits of the scalar approximation. The scalar model improves lowest
fixed-sector energies and first fixed-sector spacings over conventional LLL projection, while the tested perturbative models
give more accurate lowest energies. With the same three-body term retained, the hybrid reduces the lowest-energy errors relative
to the full second-order comparators throughout the sampled grid. Complementary five- and six-electron tests use
the isolated-pair radial amplitudes to generate explicit antisymmetric multi-LL variational states. These variational states
improve on equal-rank first-order-pair controls in energy and wavefunction diagnostics. Thus channel energies and radial amplitudes
provide distinct, quantitatively tested uses of exact pair information; the variational results are assessed with the physical
Hamiltonian rather than with the scalar or hybrid effective Hamiltonians.

These tests concern LL mixing in an ideal planar Coulomb system. Finite transverse extent modifies the projected interaction
through a layer form factor\cite{Peterson2008}, and subband mixing can also contribute in finite-width systems\cite{Peterson2013}.
Anisotropic interactions can be treated within the generalized-pseudopotential framework\cite{Yang2017}.
Such effects are separate from the radial LL-mixing response examined here. ZnO/MgZnO heterostructures motivate attention
to appreciable mixing\cite{Falson2015,Falson2018}; quantitative material predictions additionally require the relevant thickness,
screening, spin, and device parameters. By testing the two forms of pair input in explicit finite systems, this work provides
a starting point for developing a systematic microscopic description that combines nonperturbative pair correlations
with collective many-electron dynamics and induced interactions.

The present paper is organized as follows. Section~\ref{sec:two} defines the renormalized pseudopotentials, radial functions,
and scalar many-electron embedding. Section~\ref{sec:channels} analyzes the channel corrections and their perturbative limit.
Section~\ref{sec:benchmarks} presents the matched effective-Hamiltonian comparisons, and Sec.~\ref{sec:variational} tests
the radial input in variational states through six electrons. Sections~\ref{sec:discussion} and \ref{sec:conclusion}
discuss the implications and summarize the results. Appendices~\ref{app:methods}--\ref{app:variational} document
numerical methods, validation, and complete comparison tables.

\section{Two-electron spectrum and renormalized pseudopotentials}\label{sec:two}
\subsection{Hamiltonian and basis}
We consider two electrons in a plane with a perpendicular field $\mathbf B=B\hat{\mathbf z}$. We use SI units for
dimensional quantities, with $\omega_c=eB/m^*$. Lengths are measured in $\ell_B=\sqrt{\hbar/(eB)}$ and energies in $\hbar\omega_c$.
The physical Coulomb interaction is $e^2/(4\pi\epsilon_0\epsilon_r r_{12}^{\rm phys})$, where $r_{12}^{\rm phys}$ is the
dimensional electron separation. The effective Bohr radius is
\begin{equation}
 a_B^*=\frac{4\pi\epsilon_0\epsilon_r\hbar^2}{m^*e^2},
\end{equation}
and the dimensionless coupling satisfies
\begin{equation}
 \gamma_B\equiv\frac{\ell_B}{a_B^*}
 =\frac{e^2/(4\pi\epsilon_0\epsilon_r\ell_B)}{\hbar\omega_c}
 \equiv\kappa.
 \label{eq:coupling}
\end{equation}
Here $m^*$ is the effective electron mass, $\epsilon_0$ is the vacuum permittivity, and $\epsilon_r$ is the relative dielectric
constant of the material. Thus the length ratio $\gamma_B$ used throughout this work is exactly the conventional LL-mixing
energy ratio $\kappa$ for this model. Below, all coordinates and separations are dimensionless, in units of $\ell_B$.
In symmetric gauge the dimensionless Hamiltonian is
\begin{equation}
 H=\sum_{i=1}^{2}\frac12\left(-i\boldsymbol\nabla_i+\tfrac12\hat{\mathbf z}\times\mathbf r_i\right)^2 +
 \frac{\gamma_B}{|\mathbf r_2-\mathbf r_1|}.
 \label{eq:twoH}
\end{equation}
For $\mathbf R=(\mathbf r_1+\mathbf r_2)/\sqrt2$ and $\mathbf r=(\mathbf r_2-\mathbf r_1)/\sqrt2$,
\begin{align}
 H&=H_{\rm cm}+H_{\rm rel},\\
 H_{\rm cm}&=\tfrac12(-i\boldsymbol\nabla_R+\tfrac12\hat{\mathbf z}\times\mathbf R)^2,\\
 H_{\rm rel}&=\tfrac12(-i\boldsymbol\nabla_r+\tfrac12\hat{\mathbf z}\times\mathbf r)^2+\frac{\gamma_B}{\sqrt2r}.
 \label{eq:relative}
\end{align}
The scaled relative radius $r$ differs from the dimensionless electron separation $r_{12}=\sqrt2r$.

Let $\alpha=0,1,\ldots$ label the interacting radial eigenstates at fixed signed relative angular momentum $m$. To avoid confusing
the eigenstate label with the basis index, write
\begin{equation}
 \psi^{\rm rel}_{\alpha m}(r,\theta)=\frac{e^{im\theta}}{\sqrt{2\pi}}
 \sum_{n=0}^{\infty}a^{(m)}_{\alpha n}R^{(0)}_{nm}(r),
 \qquad \sum_n|a^{(m)}_{\alpha n}|^2=1.
 \label{eq:expansion}
\end{equation}
The Landau radial functions are normalized with measure $r\dd r$. A phase convention is
\begin{equation}
\begin{aligned}
 R^{(0)}_{nm}(r)={}&\sqrt{\frac{n!}{\Gamma(n+|m|+1)}}
 \left(\frac{r^2}{2}\right)^{|m|/2}\\
 &{}\times e^{-r^2/4}L_n^{|m|}(r^2/2).
\end{aligned}
\end{equation}
The noninteracting energy is $E_L(n,m)=n+(m+|m|+1)/2$. The matrix eigenproblem is
\begin{equation}
 \sum_{n'=0}^{\infty}\left[E_L(n,m)\delta_{nn'}+
 \gamma_B M^{|m|}_{nn'}\right]a^{(m)}_{\alpha n'}
 =E^{\rm rel}_{\alpha m}a^{(m)}_{\alpha n},
 \label{eq:matrix}
\end{equation}
where $M^{|m|}_{nn'}=\frac1{\sqrt2}\int_0^{\infty}\dd r\,R^{(0)}_{nm}(r)R^{(0)}_{n'm}(r)$ is the matrix of $1/(\sqrt2r)$,
following the convention of Ref.~\cite{Hai2024}. A phase change of a basis function changes matrix entries and coefficients
consistently, without changing the spectrum.

The same radial state can be expressed in terms of the correlation-resolved function introduced in Ref.~\cite{Hai2024}.
With $x=r^2/2=r_{12}^2/4$, define
\begin{equation}
\begin{aligned}
 R^{\rm rel}_{\alpha m}(r)={}&
 \sqrt{\frac{\alpha!}{\Gamma(\alpha+|m|+1)}}\,
 x^{|m|/2}e^{-x/2}h_{\alpha}^{|m|}(x),
\end{aligned}
 \label{eq:radial-h}
\end{equation}
where $R^{\rm rel}_{\alpha m}=\sum_n a^{(m)}_{\alpha n}R^{(0)}_{nm}$ and therefore
\begin{equation}
 h_{\alpha}^{|m|}(x)=\sum_{n=0}^{\infty}a^{(m)}_{\alpha n}
 \sqrt{\frac{n!\,\Gamma(\alpha+|m|+1)}
 {\alpha!\,\Gamma(n+|m|+1)}}\,L_n^{|m|}(x).
 \label{eq:h-expansion}
\end{equation}
The label $\alpha$ corresponds to the interacting radial-branch label $n$ in Ref.~\cite{Hai2024}, while $n$ here labels
the Landau basis. Choosing phases continuously from zero coupling gives $h_{\alpha}^{|m|}(x)\to L_{\alpha}^{|m|}(x)$
as $\gamma_B \to 0$. At finite coupling this function describes the Coulomb-induced radial restructuring beyond
the angular-momentum factor and Gaussian.

In our sign convention the LLL channels have $m\leq0$. Odd $|m|$ applies to fully spin-polarized electrons;
the even channels shown in the two-electron figures describe the corresponding spatially symmetric sector and
are not included in the polarized many-electron tests. The terminology correlated rotating electron pairs (CREP)
refers to these interacting relative states\cite{Hai2024}.

\subsection{Renormalized pseudopotentials and operator interpretation}
Keeping only $n=n'=0$ gives the conventional LLL Haldane pseudopotential. Direct integration of the normalized radial
function gives the standard result\cite{Haldane1983,Sodemann2013}
\begin{equation}
\begin{aligned}
 M^{|m|}_{00}&=\frac{\Gamma(|m|+1/2)}{2\,\Gamma(|m|+1)},\\
 V_{|m|}&=\gamma_B M^{|m|}_{00}
 =\frac{\gamma_B\Gamma(|m|+1/2)}{2\Gamma(|m|+1)}.
\end{aligned}
 \label{eq:lllV}
\end{equation}
The strict LLL restriction therefore gives $E^{\rm rel,LLL}_{0m}=1/2+V_{|m|}$ for $m\leq0$. It fixes $h_0^{|m|}(x)=1$,
whereas the full lowest interacting branch permits radial relaxation through higher-LL admixture. At fixed $\gamma_B$,
$V_{|m|}\sim\gamma_B/(2\sqrt{|m|})$ as $|m|\to\infty$, reflecting the increasing pair separation.

We define the \emph{renormalized pseudopotential} in channel $|m|$ from the full lowest interacting relative branch,
for $m\leq 0$, by
\begin{equation}
 V^*_{|m|}(\gamma_B)\equiv E^{\rm rel}_{0,-|m|}(\gamma_B)-\tfrac12,
 \qquad \Delta_{|m|}\equiv V^*_{|m|}-V_{|m|}.
 \label{eq:starV}
\end{equation}
This spectral definition reproduces the isolated pair channel energy relative to the noninteracting LLL kinetic reference,
while retaining radial LL mixing nonperturbatively. The quantity $\Delta_{|m|}$ measures the resulting correction to $V_{|m|}$.
The variational principle implies $\Delta_{|m|}\leq0$: the LLL radial state is an allowed trial state for the full pair problem.
Let $T_{\rm rel}$ denote the magnetic kinetic operator in Eq.~\eqref{eq:relative} and $V_C=\gamma_B/(\sqrt2r)$.
Expectations in the normalized lowest interacting state give
\begin{equation}
 V^*_{|m|}=\langle V_C\rangle_{0m}
 +\langle T_{\rm rel}\rangle_{0m}-\tfrac12,
\end{equation}
and hence
\begin{equation}
\begin{aligned}
 \Delta_{|m|}={}&\underbrace{\bigl(\langle T_{\rm rel}\rangle_{0m}-\tfrac12\bigr)}_{\geq0}\\
 &{}+\underbrace{\bigl(\langle V_C\rangle_{0m}-V_{|m|}\bigr)}_{\leq0}.
\end{aligned}
 \label{eq:relaxation-energy}
\end{equation}
For $m\leq0$, the kinetic spectrum is bounded below by $1/2$. Together with $\Delta_{|m|}\leq0$, this implies that the Coulomb-energy reduction compensates the kinetic-energy increase. Thus $V^*_{|m|}$ incorporates both contributions; it is neither a bare LLL Coulomb matrix element nor solely the Coulomb expectation in the interacting state.

\begin{figure}[t]\centering
\includegraphics[width=.95\linewidth]{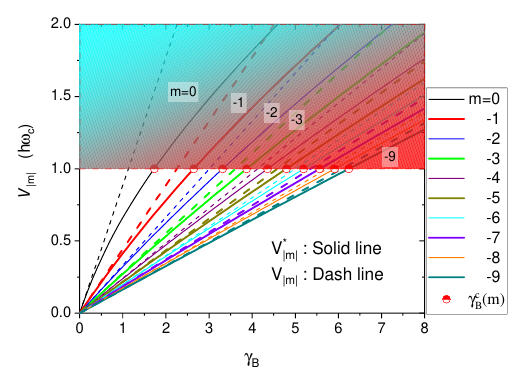}
\caption{Renormalized pseudopotentials $V^*_{|m|}$ (solid lines) and conventional Haldane pseudopotentials $V_{|m|}$ (dashed lines)
as functions of the interaction strength $\gamma_B$, for $m=0,-1,\ldots,-9$. Curves of the same color correspond to
the same angular-momentum channel. The shaded region corresponds to pseudopotential values above unity, equivalent
to $E^{\rm rel}_{0m}>3/2$ for the interacting branch. Red semi-dots identify $\gamma_B^c(m)=\sqrt{4|m|+3}$,
where $V^*_{|m|}=1$.
These mark an isolated-pair energy reference, whose interpretation is discussed in Sec.~\ref{sec:channels}.}
\label{fig:potentials}\end{figure}

Fig.~\ref{fig:potentials} compares the renormalized pseudopotentials $V^*_{|m|}$ with the conventional LLL pseudopotentials $V_{|m|}$, illustrating the energy lowering associated with interaction-induced radial relaxation.

The conventional many-electron projected interaction has the operator form
\begin{equation}
 H_{\rm LLL}=\frac N2+\sum_{i<j}\sum_{|m|=1,3,\ldots}V_{|m|}\Pi^{(ij)}_{|m|}.
\end{equation}
Here $\Pi^{(ij)}_{|m|}$ projects particles $i,j$ onto relative angular momentum $m=-|m|$ within the LLL and acts as the identity on the pair CM and spectator degrees of freedom. Its expectation value gives the probability that the selected pair occupies this relative-angular-momentum channel within the LLL.

We define the scalar renormalized-pseudopotential Hamiltonian by retaining these conventional LLL pair projectors and replacing their coefficients by the renormalized pseudopotentials:
\begin{equation}
 H_{\mathrm{ren}} =\frac N2+\sum_{i<j}\sum_{|m|=1,3,\ldots}V^*_{|m|}\Pi^{(ij)}_{|m|}.
 \label{eq:Hren}
\end{equation}
The coefficients $V^*_{|m|}$ reproduce the isolated lowest-branch pair energies by construction. In Eq.~\eqref{eq:Hren}, the effects of the interacting radial functions $h_0^{|m|}$ enter through these coefficients, while the pair projectors retain their conventional LLL form. Every projector acts in the same antisymmetric $N$-electron Hilbert space, and the sum includes all overlapping pairs. Thus, the many-electron eigenstates retain fermionic exchange and collective configuration mixing, without assigning permanent partners or factorizing the state into independent pairs.

This construction incorporates isolated-pair radial relaxation into a two-body effective interaction. Eliminating higher LLs from the full many-electron problem also generates induced three-body and higher-body operators, which are absent from $H_{\mathrm{ren}}$. Its accuracy therefore depends on how well the retained pair contribution describes the systems and observables considered; its eigenvalues need not be variational upper bounds to the physical energies. Physical observables in the effective space likewise require a consistent transformation. Section~\ref{sec:channels} examines the channel corrections and their perturbative limit, and Sec.~\ref{sec:benchmarks} tests their many-electron application against explicit multi-LL benchmarks.

\section{Channel corrections and their physical scope}\label{sec:channels}
\begin{figure}[b]\centering
\includegraphics[width=.78\linewidth]{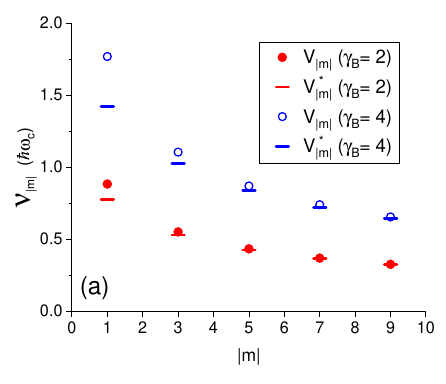}\\[4pt]
\includegraphics[width=.78\linewidth]{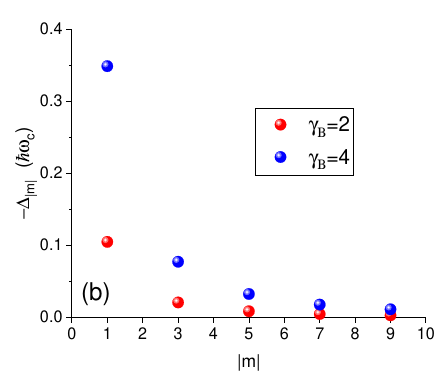}
\caption{Dependence on relative angular momentum for $m=-1,-3,-5,-7$, and $-9$ at $\gamma_B=2$ and 4: (a) conventional Haldane
pseudopotentials $V_{|m|}$ and renormalized pseudopotentials $V^*_{|m|}$; (b) the magnitude of their difference,
$-\Delta_{|m|}=V_{|m|}-V^*_{|m|}$.}
\label{fig:channels}\end{figure}

The two-electron solution shows how radial relaxation changes the interaction hierarchy among relative-angular-momentum channels.
Fig.~\ref{fig:channels} compares the conventional and renormalized pseudopotentials for odd $|m|$ at $\gamma_B=2$ and 4.
The energy lowering $-\Delta_{|m|}$ is largest in the short-range channels and decreases with increasing $|m|$, as the characteristic
pair separation increases. The stronger-coupling results show larger corrections. Microscopically, the off-diagonal matrix elements
in Eq.~\eqref{eq:matrix} admix higher radial Landau states and modify $h_0^{|m|}(x)$.

\begin{figure}[!t]\centering
\includegraphics[width=.95\linewidth]{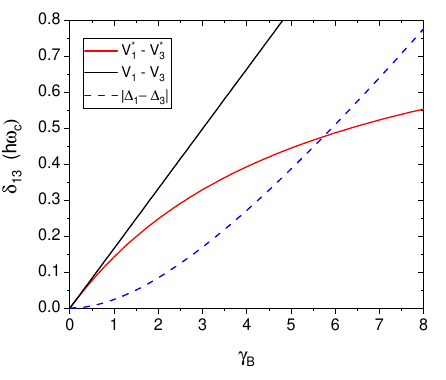}
\caption{Short-range pseudopotential differences as functions of $\gamma_B$: $\delta_{13}^{\rm LLL}=V_1-V_3$ (black solid line), $\delta_{13}^*=V_1^*-V_3^*$ (red solid line), and the magnitude of the renormalization $|\Delta_1-\Delta_3|$ (blue dashed line).}
\label{fig:difference}\end{figure}

The resulting change in the short-range hierarchy is illustrated in Fig.~\ref{fig:difference} by $\delta_{13}^{\rm LLL}=V_1-V_3$ and $\delta_{13}^*=V_1^*-V_3^*$. At $\gamma_B=3.5$, these differences are $0.58158642$ and $0.36267150$, respectively, a reduction
of $37.64\%$. Thus radial relaxation appreciably changes the relative energetic cost of the two shortest-range polarized channels,
which are relevant to Laughlin-type correlations.

For weak coupling, nondegenerate perturbation theory within a fixed $m\leq0$ channel gives
\begin{equation}
\begin{aligned}
 V^*_{|m|}&=\gamma_B M^{|m|}_{00}+\gamma_B^2 b_{|m|}
 +O(\gamma_B^3),\\
 b_{|m|}&=-\sum_{n=1}^{\infty}\frac{|M^{|m|}_{0n}|^2}{n}.
\end{aligned}
 \label{eq:pairpert}
\end{equation}
The denominator $n$ is the excitation energy of the intermediate radial Landau state relative to the LLL state in cyclotron units. The coefficient $b_{|m|}$ describes virtual excitation of the relative motion out of the LLL and its return, with the matrix elements defined in Sec.~\ref{sec:two}. In these conventions it corresponds to the two-body coefficient $V_2(m)$ of Ref.~\cite{Sodemann2013}, whose nonnegative channel index denotes our $|m|$, so that $\Delta_{|m|}=\gamma_B^2V_2(m)+O(\gamma_B^3)$\cite{Sodemann2013,Simon2013}. The $\gamma_B^2$ contribution becomes first order in $\kappa=\gamma_B$ when interaction energies are expressed in Coulomb units instead; all energies here remain in $\hbar\omega_c$.

Eq.~\eqref{eq:pairpert} therefore connects the renormalized pseudopotentials directly to the leading perturbative two-body correction. The full interacting eigenproblem retains this contribution together with higher-order radial admixture, without truncating the pair energy at a finite order in $\gamma_B$. Both descriptions use the same Coulomb interaction and the same channel; their difference is the treatment of its radial LL mixing.

\begin{figure}[htb]\centering
\includegraphics[width=.95\linewidth]{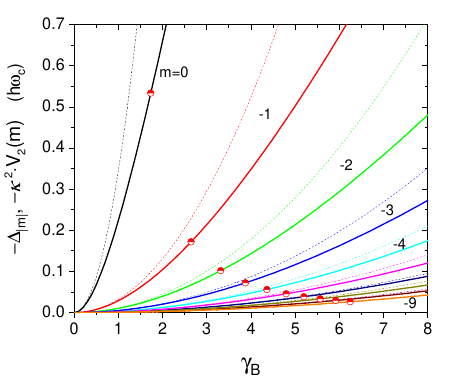}
\caption{Magnitudes of the nonperturbative correction $-\Delta_{|m|}$ (solid lines) and the quadratic two-body correction
$-\kappa^2V_2(m)$ of Ref.~\cite{Sodemann2013} (dashed lines), as functions of $\gamma_B$ for $m=0,-1,\ldots,-9$.
Curves of the same color correspond to the same channel. Here $V_2(m)=b_{|m|}$ and $\kappa=\gamma_B$.
Red semi-dots indicate the reference values $\gamma_B^c(m)=\sqrt{4|m|+3}$, at which $V^*_{|m|}=1$.}
\label{fig:pertpair}\end{figure}

Fig.~\ref{fig:pertpair} compares the resulting nonperturbative correction with its quadratic approximation. Their agreement at weak coupling confirms the expected perturbative limit. As $\gamma_B$ increases, the quadratic approximation predicts a larger magnitude of the energy lowering than the full pair solution, with the largest absolute discrepancies occurring in the short-range channels. The growing separation between the curves quantifies the importance of contributions beyond the leading perturbative correction to the isolated-pair energies.

The marked reference values in Figs.~\ref{fig:potentials} and \ref{fig:pertpair} follow from particular analytical solutions\cite{Taut1994,Hai2024}: for $m\leq0$, $E^{\rm rel}_{0m}=3/2$ at $\gamma_B^c(m)=\sqrt{4|m|+3}$. This equality identifies a pair-energy reference. LL admixture is already present below it, and coincident energies in distinct conserved angular-momentum sectors do not imply hybridization. Accordingly, $\gamma_B^c(m)$ is not a universal boundary of validity of an LLL effective description.

The coupling range has a useful material scale. For illustration, the parabolic-band parameters $m^*=0.29m_e$ (with $m_e$ the free-electron mass) and $\epsilon_r=8.5$, as tabulated in Ref.~\cite{Sodemann2013}, give $\gamma_B\simeq16.7/\sqrt{B({\rm T})}$; fields of roughly $17$--$31$ T therefore sample $\gamma_B\simeq4$--$3$. This motivates examining strong coupling in connection with ZnO/MgZnO\cite{Falson2015,Falson2018}. This range includes $\gamma_B=3.5$, used in the many-electron benchmarks below, and therefore provides a material-motivated scale at which to assess the usefulness of the nonperturbative pair input. This estimate uses the ideal planar Coulomb model, without a heterostructure form factor or screening model.

The comparison establishes the nonperturbative change in the two-body interaction input. Neither $\delta_{13}^*$ nor its change is a bulk excitation gap. The perturbative many-electron treatment also generates induced three-body terms, which are not determined by the isolated two-electron spectrum. Thus, the improved description of pair energies motivates a direct test of their many-electron usefulness. Section~\ref{sec:benchmarks} compares the scalar and hybrid constructions with perturbative effective Hamiltonians and explicit multi-LL benchmarks under matched conditions.

\section{Matched many-electron benchmarks}\label{sec:benchmarks}
We now test whether the renormalized pseudopotentials improve finite-system many-electron predictions, first within the scalar two-body model and then with the same perturbative three-body contribution retained. All comparisons use matched physical systems and a common energy convention.
\subsection{Physical systems and comparison protocol}\label{sec:protocol}
We consider fully spin-polarized planar Coulomb systems with $N=3$ and 4, at signed total angular momentum
\begin{equation}
 L_z=-J\hbar,\qquad J=\frac{3N(N-1)}2.
\end{equation}
Explicitly, $J=9$ for $N=3$ and $J=18$ for $N=4$. These are the angular-momentum sectors of the one-third Laughlin sequence in the present convention. No scalar confinement, neutralizing background, finite thickness, disorder, or explicit Zeeman term is included. The calculations concern lowest energies and ordered low-lying levels within a finite angular-momentum sector, rather than an unconstrained planar ground state or a thermodynamic filling fraction.

Using $g$ as shorthand for $\gamma_B=\kappa$, the physical Hamiltonian is
\begin{equation}
 H_N=H_0+gV,\quad H_0=\sum_i(a_i+\tfrac12),\quad
 V=\sum_{i<j}\frac{1}{r_{ij}},
 \label{eq:manyH}
\end{equation}
with dimensionless electron separations $r_{ij}=|\mathbf r_i-\mathbf r_j|$, distinct from the scaled relative radius $r=r_{ij}/\sqrt2$ of Sec.~\ref{sec:two}, and nonnegative cyclotron quantum numbers $a_i$. The expression for $H_0$ specifies its diagonal action on a determinant with these quantum numbers. The multi-LL determinant basis satisfies $\sum_i a_i\leq K$. The dimensionless cutoff $K$ bounds the excess cyclotron energy by $K\hbar\omega_c$; it is not a single-particle LL cutoff applied independently to every electron.
We use $K=10$ and 12 for $N=3$ and $K=6$ and 8 for $N=4$.

Diagonalization of $H_N$ in this truncated basis provides the finite-cutoff multi-LL benchmark, denoted $E_{{\rm full},K}$.
All descriptions use the same geometry, $N,J,g$, and zero of energy. The six couplings $g=$0.1, 0.3, 0.5, 1.0, 2.0, and 3.5
cover weak and strong mixing. Let $P$ denote the orthogonal projector onto the many-electron LLL subspace in the chosen
angular-momentum sector, where every electron occupies the LLL. ``Conventional LLL'' means $P H_NP$, with no LL-mixing
correction. The scalar model in Eq.~\eqref{eq:Hren} supplies a direct use of the renormalized pseudopotentials
in LLL exact diagonalization. No coefficient is fitted to the many-electron benchmarks.

\subsection{Perturbative comparators and the hybrid extension}\label{sec:hybrid}
With $P$ defined above, let $Q_K$ project onto states in the same sector that contain at least one higher-LL excitation
and satisfy the total excitation cutoff $K$. Thus $PQ_K=0$. The coefficient of the full second-order LL-mixing correction is
\begin{equation}
 B_K=-PVQ_K[Q_K(H_0-N/2)Q_K]^{-1}Q_KVP.
 \label{eq:BK}
\end{equation}
The inverse is restricted to the range of $Q_K$, where $H_0-N/2$ has strictly positive eigenvalues. Direct evaluation of this
operator retains both same-pair and overlapping-pair virtual processes. It implements the second-order effective-Hamiltonian
construction that contains two-body and induced three-body terms\cite{Bishara2009,Peterson2013,Sodemann2013,Simon2013};
it is not replaced by its two-body part.

For the perturbative expressions below, we denote the unit-coupling Coulomb matrix elements by
$v^{(|m|)}_{nn\prime}\equiv M^{|m|}_{nn\prime}$. The two-body contribution uses the pair coefficients
$b_{|m|}(K)=-\sum_{n=1}^{K}|v^{(|m|)}_{0n}|^2/n$. The embedded two-body operator is
\begin{equation}
 B_{2,K}=\sum_{i<j}\sum_{|m|=1,3,\ldots}
 b_{|m|}(K)\,\Pi^{(ij)}_{|m|}.
 \label{eq:B2embed}
\end{equation}
We define the connected second-order remainder by
\begin{equation}
 B_{3,K}=B_K-B_{2,K}.
 \label{eq:subtract}
\end{equation}
Here $B_{3,K}$ is the induced three-body contribution remaining after subtraction of the embedded two-body operator.
The common cutoff is important. Subtracting $B_{2,\infty}$ directly from finite-$K$ $B_K$ would incorrectly assign
a two-body truncation remainder to the three-body contribution.

We retain the finite-$K$ perturbative comparator
\begin{equation}
 H_{23,K}^{(2)}=\frac N2P+gPVP+g^2B_K,
\end{equation}
and use as the primary comparison the tail-completed operator
\begin{equation}
 H_{23}^{(2)}=\frac N2P+gPVP+g^2(B_{2,\infty}+B_{3,K}).
 \label{eq:pert-tail}
\end{equation}
Here tail completion restores the omitted two-body radial sum; $B_{3,K}$ retains its finite-$K$ dependence.
The notation $B_{2,\infty}$ denotes the infinite-sum target, evaluated numerically using 20,000 radial terms
as specified in Appendix~\ref{app:methods}. The two-body-only perturbative model is
\begin{equation}
 H_2^{(2)}=\frac N2P+gPVP+g^2B_{2,\infty}.
\end{equation}
The subscript $2$ denotes two-body terms, $23$ denotes two- plus three-body terms, and the superscript $(2)$ specifies
the perturbative order. We diagonalize these truncated effective operators within the LLL subspace. Their eigenvalues
are not interchangeable with a strict order-by-order expansion of individual levels; diagonalization retains additional
dependence on $g$ beyond second order.

The hybrid extension is
\begin{equation}
 H_{\rm hyb}=H_{\mathrm{ren}}+g^2B_{3,K}.
 \label{eq:hybrid}
\end{equation}
It uses the same three-body term as Eq.~\eqref{eq:pert-tail}, while replacing the perturbative two-body input by $V^*_{|m|}$.
Thus this comparison isolates the benefit of the nonperturbative two-body coefficients without changing the three-body term
or mixing two-body tails. The hybrid agrees with the tail-completed effective Hamiltonian through second order and differs
in selected higher-order two-body contributions. The three-body term remains perturbative; the accuracy of the hybrid beyond second order is assessed numerically below.

\subsection{Lowest energies and individual levels}\label{sec:energies}
For each model, the lowest-energy absolute error is $\varepsilon_E=|E_{\rm model}-E_{{\rm full},K}|$. In this section,
$E_{\rm full}$ abbreviates the benchmark at the stated cutoff, and $E_{\mathrm{ren}}$ denotes the lowest eigenvalue
of $H_{\mathrm{ren}}$. The scalar $H_{\mathrm{ren}}$ improves the lowest fixed-sector energy over conventional LLL projection
in all twelve size/coupling combinations. The fraction of conventional LLL energy error removed,
\begin{equation}
 \eta_E=1-\frac{|E_{\mathrm{ren}}-E_{\rm full}|}{|E_{\rm LLL}-E_{\rm full}|},
\end{equation}
is $71.97$--$74.26\%$ for $N=3$ and $67.28$--$69.07\%$ for $N=4$. This percentage concerns LL-mixing error, not the total energy or
all correlation energy.

However, both the two-body-only perturbative model and the full second-order effective Hamiltonian give smaller lowest-energy
errors than the scalar model throughout the tested grid. Table~\ref{tab:energies} displays these comparisons. The hybrid comparison
then asks whether the renormalized pseudopotentials improve the description when the leading induced three-body contribution is
retained in both models.
\begin{table*}[!t]
\centering\small
\caption{Lowest energies in the fixed sector, in $\hbar\omega_c$, including $N/2$. The benchmark cutoffs are $K=12$ for $N=3$
and $K=8$ for $N=4$. $E_{23}^{(2)}$ denotes the tail-completed perturbative comparator in Eq.~\eqref{eq:pert-tail}.}\label{tab:energies}
\setlength{\tabcolsep}{6pt}
\begin{tabular}{rrrrrrr}\toprule
$N$ & $g$ & $E_{\rm full}$ & $E_{\mathrm{ren}}$ & $E_{2}^{(2)}$ & $E_{23}^{(2)}$ & $E_{\rm hyb}$\\\midrule
3 & 0.1 & 1.571501699 & 1.571544016 & 1.571543623 & 1.571500835 & 1.571501229\\
3 & 1.0 & 2.202008159 & 2.205977773 & 2.205587829 & 2.201198871 & 2.201605147\\
3 & 3.5 & 3.846894447 & 3.888321958 & 3.870673738 & 3.811704106 & 3.832161816\\
4 & 0.1 & 2.130315459 & 2.130399692 & 2.130399069 & 2.130313884 & 2.130314510\\
4 & 1.0 & 3.280891997 & 3.288888890 & 3.288262865 & 3.279509016 & 3.280168927\\
4 & 3.5 & 6.292923438 & 6.378632040 & 6.348187474 & 6.229014393 & 6.265645136\\
\bottomrule
\end{tabular}
\end{table*}

\begin{figure*}[!t]\centering
\includegraphics[width=.90\textwidth]{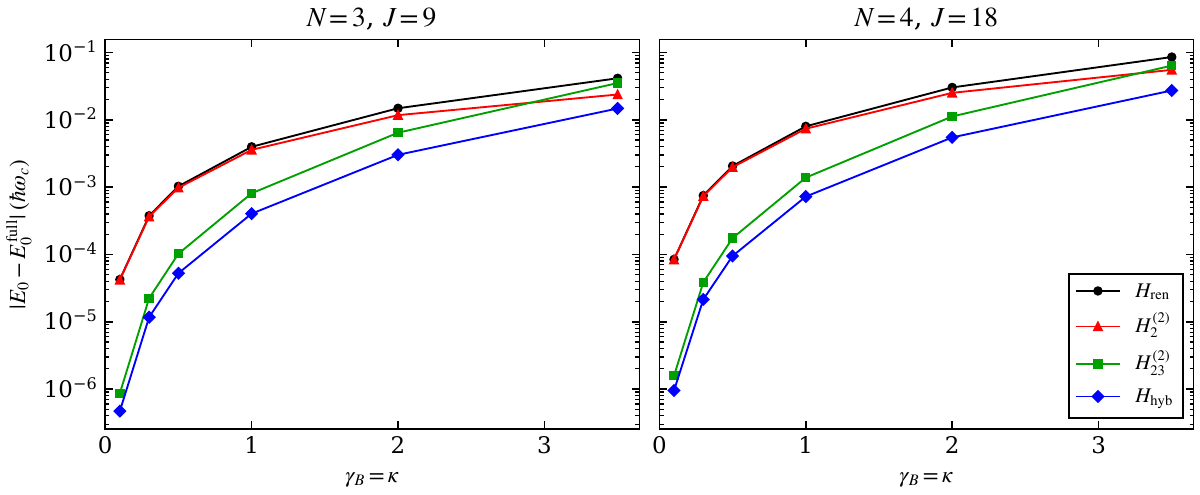}
\caption{Absolute lowest-energy errors $|E_0-E_0^{\rm full}|$ against the matched physical multi-LL benchmark
at $K=12$ for $N=3$ and $K=8$ for $N=4$. $H_{\mathrm{ren}}$ uses scalar exact-spectrum coefficients;
$H_2^{(2)}$ is two-body-only perturbation theory; $H_{23}^{(2)}$ includes the induced three-body term and
the completed two-body tail; $H_{\rm hyb}$ uses the same three-body term with exact-spectrum two-body coefficients.
Lines connect the sampled couplings. Errors are in cyclotron units, with logarithmic vertical scale.}
\label{fig:errors}\end{figure*}

The hybrid gives smaller lowest-energy errors than both $H_{23,K}^{(2)}$ and $H_{23}^{(2)}$ in every sampled case.
Fig.~\ref{fig:errors} and Appendix~\ref{app:data} show the complete scan. At $g=3.5$, the absolute hybrid errors
are $0.0147326$ and $0.0272783$ for $N=$3 and 4, compared with $0.0351903$ and $0.0639090$ for the tail-completed
perturbative comparator. These correspond to error reductions of approximately $58\%$ and $57\%$, respectively,
at a common three-body term. Both effective models can lie below the physical benchmark, consistent with their
nonvariational character.

For $g=$0.1, 1.0, and 3.5, four low-lying physical levels were recomputed at both cutoffs. At the larger cutoff,
the hybrid improves all four individual level energies relative to both full perturbative comparators in these cases.
At each coupling, levels are compared in ascending energy order within the specified sector, without tracking individual
eigenstates through possible crossings. The scalar errors vary between levels. For $N=3$ and $g=3.5$, their signs differ
across the first four levels (Appendix~\ref{app:data}). Its spectral discrepancies therefore cannot be represented
by a single additive energy shift.

\subsection{Spacings and convergence}\label{sec:spacings}
Let $E_0\leq E_1\leq\cdots$ denote the ordered levels within the fixed sector and $s_{01}=E_1-E_0$ its first spacing.
We distinguish the fraction of LLL spacing error removed,
\begin{equation}
 \eta_s=1-\frac{|s_{01}^{\rm model}-s_{01}^{\rm full}|}
 {|s_{01}^{\rm LLL}-s_{01}^{\rm full}|},
\end{equation}
from the relative spacing error $|s_{01}^{\rm model}-s_{01}^{\rm full}|/s_{01}^{\rm full}$ reported as a percentage
in Table~\ref{tab:spacings}. The recovery fraction $\eta_s$ measures improvement relative to the conventional LLL error,
whereas the relative spacing error measures the remaining discrepancy relative to the benchmark spacing itself.
For the first fixed-sector spacing, the scalar model removes approximately $95.83$--$99.53\%$ of the conventional LLL spacing
error for $N=3$ and $65.48$--$73.79\%$ for $N=4$. Thus, the scalar model improves the first spacing over LLL projection
even though the perturbative models yield better lowest energies. Full numerical values appear in Table~\ref{tab:scalar-gaps}.
\begin{table*}[!t]
\centering\small
\caption{Strong-coupling results at $g=3.5$, with benchmark cutoffs $K=12$ for $N=3$ and $K=8$ for $N=4$. Energies and spacings
are in $\hbar\omega_c$; $E_0-E_0^{\rm full}$ is a signed error. The relative spacing error uses the physical spacing as denominator.
Spectral positions are sorted within the fixed sector, including allowed center-of-mass sectors.}\label{tab:spacings}
\setlength{\tabcolsep}{6pt}
\begin{tabular}{rlrrr}\toprule
$N$ & Model & $E_0-E_0^{\rm full}$ & $E_1-E_0$ & Spacing error (\%)\\\midrule
3 & Physical benchmark & +0.00000000 & 0.310202018 & 0.000\\
3 & Scalar $H_{\mathrm{ren}}$ & +0.04142751 & 0.302868591 & 2.364\\
3 & $H_{23}^{(2)}$ & -0.03519034 & 0.211064852 & 31.959\\
3 & $H_{\rm hyb}$ & -0.01473263 & 0.300560808 & 3.108\\
4 & Physical benchmark & +0.00000000 & 0.260393592 & 0.000\\
4 & Scalar $H_{\mathrm{ren}}$ & +0.08570860 & 0.218158884 & 16.220\\
4 & $H_{23}^{(2)}$ & -0.06390904 & 0.194301438 & 25.382\\
4 & $H_{\rm hyb}$ & -0.02727830 & 0.261166642 & 0.297\\
\bottomrule
\end{tabular}
\end{table*}

At $g=3.5$, the hybrid first-spacing error is $3.108\%$ for $N=3$, slightly larger than the scalar model's $2.364\%$,
while for $N=4$ it decreases from $16.220\%$ to $0.297\%$. A spacing error is the difference between the signed errors
of its two constituent levels. Cancellation can therefore yield an accurate spacing even when the individual energies
are less accurate; reducing both individual errors need not improve their difference. For example, for $N=3$ and $g=3.5$,
the hybrid error in $E_3-E_2$ is approximately $0.03793$, compared with $0.00755$ for $H_{23}^{(2)}$. The accuracy of
individual energies and of their differences must therefore be assessed separately.

The lowest-energy improvement and the principal first-spacing conclusions persist at both physical cutoffs. At $g=3.5$,
the benchmark lowest fixed-sector energy changes by $1.35\times10^{-6}$ for $N=3$ and $1.85\times10^{-5}$ for $N=4$;
first spacings change by $1.29\times10^{-5}$ and $6.19\times10^{-5}$. These benchmark lowest-energy shifts are about
$9.2\times10^{-5}$ and $6.8\times10^{-4}$ of the corresponding hybrid energy errors, respectively, and are much smaller
than the observed hybrid advantage over the perturbative comparator. Table~\ref{tab:cutoffs} gives changes in four-level
spectra and effective models. These cutoff changes support the numerical stability of the reported energy and spacing comparisons.

The levels are ordered without restricting the center-of-mass motion to its ground state. The quoted spacings characterize
excitations within the chosen angular-momentum sector; they do not represent bulk neutral or activation gaps.

The comparison separates two ingredients of an effective many-electron description: relaxation within an interacting pair and induced
contributions involving more than two electrons. The renormalized pseudopotentials improve on LLL projection, but their isolated-pair
accuracy does not ensure a better many-electron approximation than perturbation theory. Holding the induced three-body term fixed exposes
a useful contribution from the nonperturbative pair input: the hybrid reduces the tested individual-energy errors. This provides a
quantitative basis for retaining exact pair renormalization within a broader operator expansion, with spectral accuracy assessed
observable by observable.

\section{Complementary variational tests through six electrons}\label{sec:variational}
Section~\ref{sec:benchmarks} tested the pair energies as input to many-electron effective Hamiltonians. We now examine the complementary information supplied by the same two-electron solution: its radial wavefunctions. We use these functions to construct explicit antisymmetric multi-LL trial states for larger systems and evaluate their energies with the original physical Hamiltonian $H_N$ of Sec.~\ref{sec:protocol}.

Let $c_{n,m}=a^{(m)}_{0n}$ denote the coefficients of the lowest interacting radial eigenstate in each fixed-$m$ channel in Eq.~\eqref{eq:expansion}, evaluated at the same coupling $g=\gamma_B$ as the many-electron system. These coefficients encode
the correlation-resolved radial function $h_0^{|m|}(x)$ through Eq.~\eqref{eq:h-expansion}. Writing $|n,m\rangle$ for
the noninteracting relative basis state with radial index $n$ and signed angular momentum $m$, their ratios define a radial
excitation operator, embedded with the identity on pair CM and spectator coordinates,
\begin{equation}
\begin{aligned}
 A_m^{(ij)}&=\sum_{n>0}\frac{c_{n,m}}{c_{0,m}}
 |n,m\rangle_{ij}\langle0,m|,\\
 D&=P_K\left(\sum_{i<j,m}A_m^{(ij)}\right)P_K .
\end{aligned}
 \label{eq:generator}
\end{equation}
Here $P_K$ projects onto the physical determinant space at the specified $N,J,K$. In terms of the projectors defined in
Sec.~\ref{sec:hybrid}, $P_K=P+Q_K$: it contains both the LLL subspace and the retained higher-LL states. The sum includes
all allowed odd signed relative channels, including positive $m$, subject to the cutoff imposed by $P_K$. Each $A_m^{(ij)}$
maps the $n=0$ relative radial state into states with $n>0$ at fixed $m$. For $m<0$, the initial state belongs to the LLL;
for $m>0$, it is the lowest radial state of that channel but already has nonzero cyclotron excitation. The generator $D$
is not an effective Hamiltonian and need not be Hermitian. All pairs act on one antisymmetric many-electron state.

With the lowest $s$ conventional LLL Coulomb eigenstates $\phi_a$ in the same fixed-$(N,J)$ sector as seeds, we form
\begin{equation}
 \mathcal S_s=\operatorname{span}\{D^k\phi_a:
 a=1,\ldots,s;\ k=0,\ldots,4\}.
 \label{eq:trialspace}
\end{equation}
The term $k=0$ retains the seed states, while higher powers generate additional correlated configurations. After removing linear dependencies among the $5s$ candidate vectors, Rayleigh--Ritz minimization of $H_N$ within this span determines the collective coefficients. The isolated pair input is not fitted to benchmark energies or states, but the collective coefficients are optimized separately for each many-electron system. Repeated applications of $D$ include successive corrections on different, overlapping pairs; mixing the seed configurations supplies additional collective freedom.

An equal-rank variational control replaces $c_{n,m}/c_{0,m}$ by the first-order pair amplitudes $-g v^{(|m|)}_{n0}/n$, retaining the same seeds, powers, cutoff, and minimization. Here $v^{(|m|)}_{n0}$ is the unit-coupling Coulomb matrix element defined in Sec.~\ref{sec:hybrid}, and $n$ is the excitation energy relative to the $n=0$ state at fixed $m$, in units of $\hbar\omega_c$. This control is distinct from the full second-order effective Hamiltonian of that section. It isolates the usefulness of the nonperturbative radial input within the chosen trial-space prescription.

We consider $N=5,J=30$ and $N=6,J=45$ at $g=\gamma_B=3.5$, continuing the angular-momentum sequence used in Sec.~\ref{sec:protocol}. At the larger tested cutoffs, $K=5$ and $4$, respectively, eight seeds and powers $k=0,\ldots,4$ yield 40 linearly independent vectors for each construction. Table~\ref{tab:var-main} compares their lowest fixed-sector energies. The enlarged union space is assessed separately in Appendix~\ref{app:variational}. We quantify recovery of the LL-mixing energy lowering by
\begin{equation}
 \eta_{\rm var}=\frac{E_{\rm LLL}-E_{\rm trial}}
 {E_{\rm LLL}-E_{{\rm full},K}} .
 \label{eq:recovery}
\end{equation}
Here $E_{\rm LLL}$ is the lowest conventional LLL Coulomb energy in the chosen sector, $E_{{\rm full},K}$ is the lowest physical benchmark energy at cutoff $K$, and $E_{\rm trial}$ is the lowest Rayleigh--Ritz energy in the trial space at that cutoff. The labels $E_{\rm CREP}$ and $E_{\rm control}$ distinguish the two trial constructions; $\eta_{\rm CREP}$ denotes Eq.~\eqref{eq:recovery} evaluated with $E_{\rm trial}=E_{\rm CREP}$. Because the trial space contains the lowest LLL seed, the variational principle gives $E_{{\rm full},K}\leq E_{\rm trial}\leq E_{\rm LLL}$. Thus $\eta_{\rm var}$ measures the fraction of LLL energy error removed, as does $\eta_E$ in Sec.~\ref{sec:energies}; the variational ordering makes absolute values unnecessary here.
\begin{table*}[!t]\centering\small
\caption{Complementary physical variational tests at $\gamma_B=3.5$, with eight seeds and rank 40 for both trial constructions. The residual error is $\varepsilon_{\rm trial}=E_{\rm trial}-E_{{\rm full},K}$. All energies are in $\hbar\omega_c$. Recovery is defined by Eq.~\eqref{eq:recovery}. The corresponding LLL energies are 9.595379947 and 13.010529470.}
\label{tab:var-main}
\setlength{\tabcolsep}{4pt}
\begin{tabular}{rrrrrrrr}\toprule
$N$ & $K$ & $E_{{\rm full},K}$ & $E_{\rm CREP}$ & $E_{\rm control}$ & $\varepsilon_{\rm CREP}$ & $\varepsilon_{\rm control}$ & $\eta_{\rm CREP}$\\\midrule
5 & 5 & 9.200869780 & 9.203210140 & 9.214741457 & 0.002340360 & 0.013871677 & 99.41\% \\
6 & 4 & 12.481240131 & 12.485960296 & 12.501775514 & 0.004720165 & 0.020535383 & 99.11\% \\
\bottomrule\end{tabular}\end{table*}

The CREP construction recovers $99.41\%$ and $99.11\%$ of this energy lowering for $N=5$ and $6$, respectively, and reduces the residual energy errors by factors of $5.93$ and $4.35$ relative to the equal-rank control. This improvement extends to the wavefunctions. Denoting the normalized lowest-energy benchmark and variational states in the same sector by $\Psi_{{\rm full},K}$ and $\Psi_{\rm trial}$, respectively, we define their fidelity as $\mathcal F=|\langle\Psi_{{\rm full},K}|\Psi_{\rm trial}\rangle|^2$. For $N=5$ and $6$, the CREP fidelities are $0.99790$ and $0.99509$, respectively, compared with $0.99105$ and $0.98053$ for the equal-rank control. The CREP states also reproduce the benchmark pair-channel populations more closely in both systems, as quantified by the aggregate population distance defined and tabulated in Appendix~\ref{app:variational}.

The improvement over the control persists at both tested cutoffs (Appendix~\ref{app:variational}). For six electrons, the benchmark energy changes by $0.0022435$ from $K=3$ to $4$, appreciable compared with the $K=4$ CREP residual error $0.0047202$. Eight seed eigenstates are retained from the 1,206-dimensional LLL sector for $N=6$; the seed-dependence tests are reported in Appendix~\ref{app:variational}.

The improvements in energy, full-state overlap, and aggregate pair-channel populations support the usefulness of interacting-pair radial functions combined with collective configuration mixing. These explicit multi-LL states provide variational upper bounds to the corresponding physical sector energies and extend the assessment of the pair input beyond the effective-Hamiltonian spectra of Sec.~\ref{sec:benchmarks}.

\section{Discussion}\label{sec:discussion}
The results identify two complementary ways in which exact two-electron information contributes to a many-electron description of LL mixing. The renormalized pseudopotentials retain isolated-pair radial relaxation through the pair energies, including both Coulomb- and kinetic-energy changes. The corresponding radial wavefunctions supply amplitudes for constructing explicit multi-LL trial states. In both cases, the pair input is embedded in a common antisymmetric many-electron space, where collective configuration mixing determines its usefulness.

The effective-Hamiltonian comparisons in Sec.~\ref{sec:benchmarks} show why the pair input and the induced many-body terms must be assessed together. Virtual LL excitations involving different, overlapping pairs generate contributions beyond the isolated two-body interaction. The scalar $H_{\mathrm{ren}}$ improves on conventional LLL projection, while the tested perturbative Hamiltonians give more accurate lowest fixed-sector energies. Retaining the same second-order induced three-body term in both descriptions, however, replacing the perturbative two-body input by the renormalized pseudopotentials reduces the absolute errors of these energies throughout the sampled grid. The hybrid and the tail-completed perturbative Hamiltonian agree through second order in the coupling; their comparison at stronger coupling therefore tests the additional nonperturbative pair information against a common perturbative reference. Improvements in individual energies do not automatically imply improvements in every spacing, as the examples in Sec.~\ref{sec:spacings} demonstrate.

The variational tests in Sec.~\ref{sec:variational} establish a complementary use of the radial wavefunctions. Repeated pair corrections and mixing of LLL seed configurations combine fixed isolated-pair amplitudes with collective variational freedom. For five and six electrons, this construction improves energies, full-state squared overlaps, and aggregate pair-channel populations relative to controls using first-order pair amplitudes at equal rank. These results show that the radial structure obtained from the two-electron problem remains useful when the many-electron state is allowed to reorganize collectively. Together with the effective-Hamiltonian comparisons, they support the use of both spectral and wavefunction information from the same microscopic pair solution.

These two uses of pair information also clarify the treatment of observables. Renormalized pseudopotentials specify an effective interaction but do not by themselves determine a map to physical wavefunctions. Densities and pair correlations evaluated in an effective space require consistently transformed operators or reconstructed physical states. The explicit multi-LL variational states instead permit direct evaluation of physical observables, as illustrated by the pair-channel populations. The associated numerical checks are reported in Sec.~\ref{sec:variational} and Appendix~\ref{app:variational}.

A systematic next step is to combine the renormalized pseudopotentials with nonperturbative connected few-electron operators. This requires a specified effective space and a consistent matching or decoupling map, followed by subtraction of embedded lower-body contributions. Testing those operators in larger systems with the few-electron input held fixed would assess their predictive usefulness. The present hybrid, with its perturbative three-body contribution, provides a quantitative reference for this development, while the variational construction offers a complementary route for incorporating pair radial structure into collective states.

The strong-coupling range considered here is motivated in part by ZnO/MgZnO systems, as discussed in Sec.~\ref{sec:channels}. The present calculations establish changes in the short-range pair hierarchy and their consequences for the tested finite systems within the ideal planar Coulomb model. Connecting these results to bulk gaps and phase stability requires larger-system studies; quantitative comparison with material-specific activation gaps also requires appropriate treatment of finite thickness, screening, spin, and device conditions. The results provide a starting point for such studies by identifying useful nonperturbative pair input and explicit ways to test its many-electron application.

\section{Conclusion}\label{sec:conclusion}
We have defined renormalized pseudopotentials from the exact two-electron relative-motion spectrum and tested their usefulness in finite many-electron systems. They incorporate isolated-pair LL mixing nonperturbatively, recover the leading two-body perturbative correction, and substantially modify the short-range interaction hierarchy at stronger coupling.

The three- and four-electron comparisons demonstrate the complementary roles of pair renormalization and induced many-body interactions. The scalar model improves on conventional LLL projection, while the tested perturbative models give more accurate lowest fixed-sector energies. With the same second-order induced three-body term retained, the renormalized pseudopotentials reduce the absolute lowest-energy errors relative to both full second-order comparators throughout the sampled grid. At $\gamma_B=3.5$, the reductions relative to the tail-completed two-plus-three-body comparator are approximately $58\%$ and $57\%$ for three and four electrons, respectively.

The same two-electron solution also supplies useful radial amplitudes for collective variational states. For five and six electrons, the resulting trial states recover $99.41\%$ and $99.11\%$ of the LL-mixing energy lowering relative to their matched finite-cutoff benchmarks. They outperform equal-rank first-order-pair controls in energy, full-state squared overlap, and aggregate pair-channel populations.

Together, these results establish exact interacting-pair energies and radial structure as quantitatively useful building blocks for LL-mixed many-electron descriptions. They provide a starting point for developing a systematic microscopic description of quantum Hall systems, combining nonperturbative two-electron correlations with collective many-electron dynamics and induced interactions.

\begin{acknowledgments}
This research was supported by FAPESP (S\~ao Paulo Research Foundation, grant 2024/16063-6), FAPEG, FAPEMIG, and CNPq (Brazil).

\end{acknowledgments}

\appendix
\section{Numerical definitions and validation}\label{app:methods}
\subsection{Radial and many-electron bases}
The pair calculations use a 240-function radial basis, with comparisons to 160 functions. In the phase convention of Eq.~\eqref{eq:expansion}, the unit-coupling Coulomb matrix is $v^{(\mu)}_{nn\prime}=M^{|m|}_{nn\prime}$ with $\mu=|m|$, so that
\begin{equation}
\begin{aligned}
 v^{(\mu)}_{nn'}={}&\frac12\sqrt{\frac{n!n'!}{\Gamma(n+\mu+1)\Gamma(n'+\mu+1)}}\\
 &{}\times\int_0^\infty\dd x\,x^{\mu-1/2}e^{-x}L_n^\mu(x)L_{n'}^\mu(x),\\
 &\mu=|m|.
\end{aligned}
\end{equation}
The oscillator implementation uses radial basis states multiplied by $(-1)^n$ relative to Eq.~\eqref{eq:expansion}. Accordingly, its Coulomb matrix elements carry the factor $(-1)^{n+n\prime}$ and its radial expansion coefficients carry $(-1)^n$. Using this phase convention consistently in the pair transformations and radial coefficients preserves the operator in Eq.~\eqref{eq:generator}, the spectra, and the perturbative sums in Eq.~\eqref{eq:pairpert}. The magnitude of the analytic row needed for the perturbative sum is
\begin{equation}
 |v^{(\mu)}_{0n}|=\frac{\Gamma(\mu+1/2)(1/2)_n}
 {2\sqrt{\Gamma(\mu+1)n!\Gamma(n+\mu+1)}},
\end{equation}
where $(1/2)_n=\Gamma(n+1/2)/\Gamma(1/2)$. The maximum absolute discrepancy between this expression and the magnitudes of the numerical matrix elements is below $2\times10^{-15}$ for $\mu=1,3,5$ and $n=0,\ldots,39$. The perturbative two-body sum uses 20,000 terms.

A single-particle orbital is labeled by nonnegative integers $(a,b)$, with kinetic energy $(a+1/2)\hbar\omega_c$ and
signed angular momentum $(a-b)\hbar$. The many-electron basis consists of antisymmetric Slater determinants satisfying
$\sum_i(b_i-a_i)=J$ and $\sum_i a_i\leq K$. The LLL sectors $(N,J)=(3,9)$ and $(4,18)$ have dimensions 7 and 34, respectively.
Including higher LLs expands the basis to 7,690 and 14,459 states for $N=3$ at $K=10$ and $12$, respectively, and to 20,575
and 59,968 states for $N=4$ at $K=6$ and $8$, respectively. Coulomb matrix elements are evaluated using transformations
between single-particle and relative/CM oscillator bases, with fermionic signs retained throughout. All pairwise Coulomb
terms are included, and no pair-channel populations are fitted.

The lowest physical eigenpairs are obtained by sparse diagonalization, while the effective Hamiltonians are diagonalized directly in the LLL basis. The lowest two physical levels are evaluated over the full six-coupling grid. Four-level comparisons are available at $g=0.1,1.0,3.5$; at $g=0.3,0.5,2.0$, the comparisons use the lowest fixed-sector energy and first spacing.

\subsection{Connected subtraction and small-coupling checks}
In the two-electron sector $(N,J)=(2,3)$, the connected subtraction in Eq.~\eqref{eq:subtract} vanishes to numerical precision: the matrix norms of $B_{3,K}$ are $9.5\times10^{-17}$ and $1.15\times10^{-16}$ at $K=6$ and $8$, respectively. The matrix difference between the exact two-body operator and its second-order expansion scales as $g^3$. At $g=0.02,0.01,0.005$, the residual divided by $g^3$ is approximately $0.0138225,0.0138352,0.0138413$, respectively.

The effect of completing the two-body tail is larger than the residual cutoff dependence of $B_{3,K}$. Comparing $H_{23,K}^{(2)}$ with the tail-completed Hamiltonian therefore separates two-body truncation effects from the induced three-body contribution.

\section{Complete numerical comparisons}\label{app:data}
Tables~\ref{tab:scan}--\ref{tab:cutoffs} collect the energy and spacing comparisons for three and four electrons, using the conventions of Sec.~\ref{sec:protocol}. Table~\ref{tab:scan} covers the full coupling grid, and Table~\ref{tab:scalar-gaps} gives the scalar-model first spacings. Table~\ref{tab:levels} resolves the strong-coupling errors by level, showing that the scalar discrepancy cannot be represented by a common additive shift. Table~\ref{tab:cutoffs} reports the changes under enlargement of the physical basis.
\begin{table*}[!t]
\centering\small
\caption{Complete lowest-energy scan. The last five columns give absolute lowest-energy errors $|E_0^{\rm model}-E_0^{\rm full}|$ at $K=12$ for $N=3$ and $K=8$ for $N=4$; all values are in $\hbar\omega_c$. $H_{23,K}^{(2)}$ uses the uncompleted finite-$K$ two-body tail.}\label{tab:scan}
\setlength{\tabcolsep}{4pt}
\begin{tabular}{rrrrrrrrr}\toprule
$N$ & $g$ & $E_{\rm LLL}$ & $E_{\rm full}$ & $H_{\mathrm{ren}}$ & $H_2^{(2)}$ & $H_{23,K}^{(2)}$ & $H_{23}^{(2)}$ & $H_{\rm hyb}$\\\midrule
3 & 0.1 & 1.571652665 & 1.571501699 & 4.232e-05 & 4.192e-05 & 8.123e-07 & 8.640e-07 & 4.700e-07\\
3 & 0.3 & 1.714957994 & 1.713611248 & 3.754e-04 & 3.648e-04 & 2.186e-05 & 2.235e-05 & 1.166e-05\\
3 & 0.5 & 1.858263323 & 1.854554919 & 1.028e-03 & 9.789e-04 & 1.009e-04 & 1.024e-04 & 5.251e-05\\
3 & 1.0 & 2.216526646 & 2.202008159 & 3.970e-03 & 3.580e-03 & 8.023e-04 & 8.093e-04 & 4.030e-04\\
3 & 2.0 & 2.933053293 & 2.877333948 & 1.485e-02 & 1.173e-02 & 6.393e-03 & 6.434e-03 & 3.018e-03\\
3 & 3.5 & 4.007843262 & 3.846894447 & 4.143e-02 & 2.378e-02 & 3.494e-02 & 3.519e-02 & 1.473e-02\\
4 & 0.1 & 2.130572905 & 2.130315459 & 8.423e-05 & 8.361e-05 & 1.343e-06 & 1.575e-06 & 9.490e-07\\
4 & 0.3 & 2.391718714 & 2.389420349 & 7.493e-04 & 7.324e-04 & 3.628e-05 & 3.850e-05 & 2.143e-05\\
4 & 0.5 & 2.652864523 & 2.646531169 & 2.057e-03 & 1.979e-03 & 1.681e-04 & 1.747e-04 & 9.484e-05\\
4 & 1.0 & 3.305729047 & 3.280891997 & 7.997e-03 & 7.371e-03 & 1.352e-03 & 1.383e-03 & 7.231e-04\\
4 & 2.0 & 4.611458093 & 4.515858286 & 3.027e-02 & 2.515e-02 & 1.103e-02 & 1.122e-02 & 5.474e-03\\
4 & 3.5 & 6.570051663 & 6.292923438 & 8.571e-02 & 5.526e-02 & 6.276e-02 & 6.391e-02 & 2.728e-02\\
\bottomrule
\end{tabular}
\end{table*}

\begin{table*}[!t]
\centering\small
\caption{Scalar-model first spacings $E_1-E_0$ and cutoff changes, in $\hbar\omega_c$. Full denotes the multi-LL benchmark at $K=12$ for $N=3$ and $K=8$ for $N=4$. Error removed is relative to the conventional LLL spacing error. $\delta_K$ is the physical spacing at the larger cutoff minus that at the smaller cutoff.}\label{tab:scalar-gaps}
\setlength{\tabcolsep}{6pt}
\begin{tabular}{rrrrrrr}\toprule
$N$ & $g$ & Full & LLL & Scalar & Error removed (\%) & $\delta_K$\\\midrule
3 & 0.1 & 0.013657805 & 0.013885489 & 0.013660790 & 98.689 & -5.47e-07\\
3 & 0.3 & 0.039670706 & 0.041656466 & 0.039688380 & 99.110 & -3.94e-06\\
3 & 0.5 & 0.064079956 & 0.069427443 & 0.064105173 & 99.528 & -8.73e-06\\
3 & 1.0 & 0.119019314 & 0.138854887 & 0.118912534 & 99.462 & -1.98e-05\\
3 & 2.0 & 0.208752298 & 0.277709774 & 0.207170498 & 97.706 & -2.48e-05\\
3 & 3.5 & 0.310202018 & 0.485992104 & 0.302868591 & 95.828 & -1.29e-05\\
4 & 0.1 & 0.010770628 & 0.010935050 & 0.010727540 & 73.794 & -1.83e-06\\
4 & 0.3 & 0.031374127 & 0.032805149 & 0.030990343 & 73.181 & -1.36e-05\\
4 & 0.5 & 0.050830117 & 0.054675249 & 0.049775140 & 72.563 & -3.08e-05\\
4 & 1.0 & 0.095171086 & 0.109350498 & 0.091064426 & 71.038 & -7.43e-05\\
4 & 2.0 & 0.169997216 & 0.218700995 & 0.154555624 & 68.295 & -1.05e-04\\
4 & 3.5 & 0.260393592 & 0.382726742 & 0.218158884 & 65.476 & -6.19e-05\\
\bottomrule
\end{tabular}
\end{table*}

\begin{table*}[!t]
\centering\small
\caption{Signed individual-level errors $E_i^{\rm model}-E_i^{\rm full}$ in $\hbar\omega_c$ at $g=3.5$, with $K=12$ for $N=3$ and $K=8$ for $N=4$. Levels are sorted within the fixed sector.}\label{tab:levels}
\setlength{\tabcolsep}{6pt}
\begin{tabular}{rlrrrr}\toprule
$N$ & Model & Level 0 & Level 1 & Level 2 & Level 3\\\midrule
3 & $H_{\mathrm{ren}}$ & +0.04142751 & +0.03409408 & -0.02887025 & +0.07747416\\
3 & $H_{\rm hyb}$ & -0.01473263 & -0.02437384 & -0.02972258 & +0.00820998\\
3 & $H_{23}^{(2)}$ & -0.03519034 & -0.13432751 & -0.11407028 & -0.12162296\\
4 & $H_{\mathrm{ren}}$ & +0.08570860 & +0.04347389 & +0.07671208 & +0.09072386\\
4 & $H_{\rm hyb}$ & -0.02727830 & -0.02650525 & -0.03468081 & -0.03354698\\
4 & $H_{23}^{(2)}$ & -0.06390904 & -0.13000120 & -0.14042774 & -0.12866364\\
\bottomrule
\end{tabular}
\end{table*}

\begin{table*}[!t]
\centering\small
\caption{Absolute changes in energies and spacings, in $\hbar\omega_c$, at $g=3.5$: $K=10\to12$ for $N=3$ and $K=6\to8$ for $N=4$. The maximum is over four sorted levels.}\label{tab:cutoffs}
\setlength{\tabcolsep}{6pt}
\begin{tabular}{rlrr}\toprule
$N$ & Description & Max. level change & First-spacing change\\\midrule
3 & Physical & 2.653e-05 & 1.290e-05\\
3 & Hybrid & 1.847e-07 & 1.920e-07\\
3 & Perturbative, finite $K$ & 9.134e-04 & 6.755e-04\\
3 & Perturbative, completed tail & 1.847e-07 & 1.916e-07\\
4 & Physical & 8.034e-05 & 6.187e-05\\
4 & Hybrid & 2.868e-06 & 6.064e-08\\
4 & Perturbative, finite $K$ & 2.560e-03 & 1.663e-03\\
4 & Perturbative, completed tail & 2.870e-06 & 4.500e-07\\
\bottomrule
\end{tabular}
\end{table*}

\section{Variational construction and larger-system checks}\label{app:variational}
\subsection{Basis, optimization, and observables}
The physical determinant basis follows Appendix~\ref{app:methods}. In the $(N,J)=(5,30)$ sector, the LLL dimension is 192, and the multi-LL dimensions are 58,672 and 133,074 at $K=4$ and $5$, respectively. In the $(N,J)=(6,45)$ sector, the corresponding dimensions are 1,206 in the LLL and 221,977 and 642,174 at $K=3$ and $4$, respectively. Center-of-mass motion is retained, as in Sec.~\ref{sec:benchmarks}.

The isolated radial input uses 240 basis functions. The embedding of Eq.~\eqref{eq:generator} preserves pair CM and spectator labels and includes fermionic signs through the determinant transformations. Both the CREP generator and its first-order-pair control include all allowed odd signed-$m$ channels, including positive $m$, subject to the total excitation cutoff. Singular-value decomposition retains directions with singular value above $10^{-11}$ times the largest. The physical Hamiltonian is then diagonalized in the orthonormalized span. Since $D$ raises total cyclotron excitation, powers beyond $D^3$ vanish at $K=3$; the eight-seed space then has rank 32 rather than 40.

The seeds are the lowest $s=1,2,4,8$ LLL Coulomb eigenstates. At fixed cutoff, increasing $s$ gives nested trial spaces and cannot raise the variational energy. The many-electron coefficients are optimized separately for each system using the same isolated-pair radial input and trial-space prescription.

For a normalized physical state, define the pair-channel probability averaged over all pairs and its total-variation distance from the benchmark,
\begin{equation}
 p_m=\frac{1}{\binom N2}\sum_{i<j}
 \langle\Psi|\mathcal P_m^{(ij)}|\Psi\rangle,\qquad
 d_p=\frac12\sum_m|p_m^{\rm trial}-p_m^{\rm full}|.
\end{equation}
Here $\mathcal P_m^{(ij)}$ projects onto physical relative angular momentum $m$ and sums over radial states, whereas the pair projectors in $H_{\mathrm{ren}}$ act within the LLL. The relative density matrix is obtained by tracing over pair CM and spectator coordinates. The sum includes all allowed odd signed channels, including positive $m$ when present, and satisfies $\sum_m p_m=1$. The fidelity $\mathcal F$ is defined in Sec.~\ref{sec:variational}.
Table~\ref{tab:var-diag}
summarizes the energy errors, fidelities, and pair-channel population distances for the CREP, first-order-pair, and enlarged union constructions.

\begin{table*}[!t]\centering\small
\caption{Energy and wavefunction diagnostics at $\gamma_B=3.5$, $N=5,K=5$ and $N=6,K=4$. All rows use eight seeds. Energy errors are in $\hbar\omega_c$; $d_p$ is the total-variation distance of pair-channel populations and $\mathcal F$ the squared full-state overlap. The union includes three additional independent directions and is not an equal-rank control.}
\label{tab:var-diag}
\setlength{\tabcolsep}{6pt}
\begin{tabular}{rlrrrr}\toprule
$N$ & Construction & Rank & $E_{\rm trial}-E_{{\rm full},K}$ & $d_p$ & $\mathcal F$\\\midrule
5 & CREP & 40 & 0.002340360 & 0.00320512 & 0.99789883\\
5 & First-order pair & 40 & 0.013871677 & 0.00833078 & 0.99104626\\
5 & Union & 43 & 0.001771033 & 0.00311271 & 0.99818061\\
6 & CREP & 40 & 0.004720165 & 0.01157263 & 0.99509024\\
6 & First-order pair & 40 & 0.020535383 & 0.02651178 & 0.98053136\\
6 & Union & 43 & 0.004027249 & 0.01046196 & 0.99542517\\
\bottomrule\end{tabular}\end{table*}

\subsection{Seed dependence and the enlarged union space}
Table~\ref{tab:var-seeds} shows the seed dependence at the larger cutoffs. For five electrons, the population distance increases slightly between two and four seeds even as the energy decreases.
\begin{table*}[!t]\centering\small
\caption{Seed dependence of ordinary CREP at $\gamma_B=3.5$, $N=5,K=5$ and $N=6,K=4$. Trial energies are in $\hbar\omega_c$, $\eta_{\rm var}$ is the recovered finite-cutoff LL-mixing lowering, and $d_p$ is the pair-population distance. The LLL seed spaces have dimensions 192 and 1,206, respectively.}
\label{tab:var-seeds}
\setlength{\tabcolsep}{6pt}
\begin{tabular}{rrrrrr}\toprule
$N$ & Seeds & Rank & $E_{\rm trial}$ & $\eta_{\rm var}$ & $d_p$\\\midrule
5 & 1 & 5 & 9.207061282 & 98.4306\% & 0.02477838\\
5 & 2 & 10 & 9.204242471 & 99.1451\% & 0.00750001\\
5 & 4 & 20 & 9.204169953 & 99.1635\% & 0.00784293\\
5 & 8 & 40 & 9.203210140 & 99.4068\% & 0.00320512\\
6 & 1 & 5 & 12.489915853 & 98.3609\% & 0.04739041\\
6 & 2 & 10 & 12.486981222 & 98.9153\% & 0.01618401\\
6 & 4 & 20 & 12.486979385 & 98.9157\% & 0.01608272\\
6 & 8 & 40 & 12.485960296 & 99.1082\% & 0.01157263\\
\bottomrule\end{tabular}\end{table*}

As a separate extension, we add four first-application correction vectors to $\mathcal S_s$: $D_G\phi_1$ for channel groups $G=\{-1\},\{-3\},\{-5\},\{m\leq-7\}$, where $D_G$ restricts the sum in Eq.~\eqref{eq:generator}. These four groups partition the negative odd channels. Since the initial LLL seed $\phi_1$ has only negative relative-angular-momentum channels, their sum reproduces $D\phi_1$, which already belongs to $\mathcal S_s$. The sum relation leaves at most three additional independent directions; all three are retained in the tested spaces. We call this the union space. With eight seeds, its rank is 43 at $K=4,5$ and 35 at $K=3$. The equal-rank control is compared with the original CREP space $\mathcal S_s$.

\subsection{Cutoff dependence and internal numerical checks}
\begin{table*}[!t]\centering\small
\caption{Cutoff comparisons at $\gamma_B=3.5$ and eight seeds. Benchmark energies and trial errors are in $\hbar\omega_c$. Each error is measured against the physical benchmark at the same $K$. Control denotes the first-order-pair construction; Union includes the additional channel-group directions.}
\label{tab:var-cutoffs}
\setlength{\tabcolsep}{6pt}
\begin{tabular}{rrlrrrr}\toprule
$N$ & $K$ & Construction & Rank & $E_{{\rm full},K}$ & Trial error & $\eta_{\rm var}$\\\midrule
5 & 4 & CREP & 40 & 9.201029786 & 0.002296546 & 99.4176\% \\
5 & 4 & Union & 43 & 9.201029786 & 0.001733705 & 99.5604\% \\
5 & 4 & Control & 40 & 9.201029786 & 0.012995626 & 96.7045\% \\
5 & 5 & CREP & 40 & 9.200869780 & 0.002340360 & 99.4068\% \\
5 & 5 & Union & 43 & 9.200869780 & 0.001771033 & 99.5511\% \\
5 & 5 & Control & 40 & 9.200869780 & 0.013871677 & 96.4838\% \\
6 & 3 & CREP & 32 & 12.483483628 & 0.004376576 & 99.1696\% \\
6 & 3 & Union & 35 & 12.483483628 & 0.003702294 & 99.2975\% \\
6 & 3 & Control & 32 & 12.483483628 & 0.018252146 & 96.5369\% \\
6 & 4 & CREP & 40 & 12.481240131 & 0.004720165 & 99.1082\% \\
6 & 4 & Union & 43 & 12.481240131 & 0.004027249 & 99.2391\% \\
6 & 4 & Control & 40 & 12.481240131 & 0.020535383 & 96.1202\% \\
\bottomrule\end{tabular}\end{table*}

Table~\ref{tab:var-cutoffs} compares the benchmark energies and variational errors at the two tested cutoffs.
The five-electron benchmark energy decreases by $0.0001600065$ between $K=4$ and $5$, while the six-electron energy decreases by $0.0022434972$ between $K=3$ and $4$. At both tested cutoffs, the CREP construction recovers more than $99\%$ of the LL-mixing energy lowering and retains its improvement over the equal-rank control.

The six-electron Hamiltonian is applied in the full finite determinant basis using a sparse pair-removal transform $T$, radial Coulomb blocks, and the diagonal cyclotron energy, without assembling the full Coulomb matrix. For a normalized random vector $u$ at $K=4$, the identity $T^\dagger T=15I$ is verified with $\|T^\dagger Tu-15u\|_2=6.94\times10^{-15}$, where $15=\binom{6}{2}$ counts the electron pairs and $I$ is the identity on the determinant space. The benchmark Hamiltonian residuals are $6.38\times10^{-11}$ for $N=5,K=5$ and $1.10\times10^{-10}$ for $N=6,K=4$.

\end{document}